\IfFileExists{revtex4-2.cls}{%
  \documentclass[aps,prd,reprint,superscriptaddress,nofootinbib,amsmath,amssymb]{revtex4-2}
  \newcommand{\FALLBACK}{0}%
}{%
  \documentclass[11pt,a4paper]{article}
  \newcommand{\FALLBACK}{1}%
}

\usepackage{amsmath,amssymb, bm}

\if\FALLBACK1
  \usepackage[margin=2.4cm]{geometry}
  \newcommand{\affiliation}[1]{\thanks{#1}}
  
  \newenvironment{acknowledgments}{\subsection*{Acknowledgments}}{}
  
\fi
\usepackage[T1]{fontenc}
\usepackage[utf8]{inputenc}
\usepackage{graphicx}
\usepackage{float}
\usepackage{xcolor}
\usepackage[colorlinks=true,linkcolor=blue,citecolor=blue,urlcolor=blue]{hyperref}

\newcommand{\g}{\gamma}
\newcommand{\Bx}{\Box}

\begin{document}
\bibliographystyle{apsrev4-2}

\title{Scalar dynamics in a subclass Ia mimetic DHOST bounce}

\author{Idaiane~L.~Machado}
\email[Corresponding author: ]{idaiane.machado@hotmail.com}
\affiliation{PPGCosmo, CCE - Universidade Federal do Esp\'{\i}rito Santo, \\ Avenida Fernando Ferrari 514, 29075-910, Vitória, Esp\'{\i}rito Santo, Brazil}
\affiliation{Department of Astronomy, School of Physical Sciences, University of Science and Technology of China, Hefei 230026, China}
\affiliation{CAS Key Laboratory for Research in Galaxies and Cosmology, School of Astronomy and Space Science, 
University of Science and Technology of China, Hefei 230026, China}

\author{Yi-Fu~Cai}
\affiliation{Department of Astronomy, School of Physical Sciences, University of Science and Technology of China, Hefei 230026, China}
\affiliation{CAS Key Laboratory for Research in Galaxies and Cosmology, School of Astronomy and Space Science, 
University of Science and Technology of China, Hefei 230026, China}


\begin{abstract}

 We study scalar perturbations in a quadratic subclass Ia degenerate higher-order scalar-tensor (DHOST) realization of a nonsingular bouncing cosmology previously analyzed in higher-derivative mimetic gravity. After a constant rescaling of the scalar potential, the subclass Ia model reproduces the same scale-factor history as the original model, although the associated mimetic Lagrange multiplier generally differs. For the comparison with the original mimetic model, we focus on the branch $0<\gamma<2/3$, for which the scalar sound speed of the original model is positive. We perform an ADM and Hamiltonian reduction of the scalar sector and show that, imposing the mimetic and momentum constraints, the reduced canonical scalar mode contains no spatial-gradient contribution. The physical scalar sound speed therefore vanishes, $c_s^2=0$, in agreement with the known vacuum behavior of the \(\alpha_L=0\) mimetic-DHOST sector. On the matched bouncing background, the resulting effective frequency squared is independent of the comoving wavenumber and approaches zero from below in the remote contracting phase.
 Consequently, the asymptotic oscillatory regime used in the previous mimetic model to impose positive-frequency initial conditions is absent. Although canonical normalization can still be imposed through the Wronskian condition, it does not select the initial state uniquely. Without an additional state prescription, the asymptotic dynamics therefore does not uniquely determine the primordial scalar spectrum.

 \vspace{0.2cm}
\end{abstract}

\maketitle

\section{Introduction}
\label{sec:intro}

 Nonsingular bouncing cosmologies offer an alternative description of the early Universe in which a contracting phase is joined smoothly to an expanding branch without encountering a curvature singularity \cite{novello2008bouncing, cai2014exploring,battefeld2015critical, cai2017darkmatter,brandenberger2017bouncing}. A nonsingular background alone, however, is not sufficient to characterize the model: the dynamics of cosmological perturbations, the propagating degrees of freedom, and the prescription adopted for their initial quantum state must all be examined before the primordial spectrum can be predicted. Quantum fluctuations generated during a contracting phase can lead to nearly scale-invariant spectra for suitable backgrounds and initial conditions \cite{wands1999duality,finelli2002generation, cai2009leewick, cai2009ng,cai2012nonsingular}.
 
 Mimetic gravity provides a simple framework in which suitable
 scalar-field potentials generate nonsingular cosmological backgrounds
 \cite{chamseddine2013mimetic,chamseddine2014cosmology,
 sebastiani2017mimetic,malaeb2026theoretical}. Related nonsingular and cyclic cosmologies based on Lagrange-multiplier constructions were also explored before the development of mimetic gravity \cite{cai2011lagrange}. Higher-derivative extensions of mimetic gravity and their cosmological perturbations have also been investigated
 \cite{mirzagholi2015imperfect,arroja2016cosmological}.
 In Ref.~\cite{machado2025quantum}, scalar perturbations were analyzed for a class of bouncing solutions of a mimetic model containing a higher-derivative term proportional to $(\Box\phi)^2$. In the remote contracting phase, the corresponding canonical scalar modes enter an asymptotic oscillatory regime, which was used to impose positive-frequency initial conditions before evolving the perturbations through the contraction and the bounce. However, the minimal higher-derivative model of Ref.~\cite{machado2025quantum} should be regarded as a low-energy effective description. Hamiltonian analyses have shown that its reduced scalar sector contains an ordinary kinetic ghost, distinct from an additional Ostrogradsky degree of freedom, whose propagation is prevented by the mimetic constraint \cite{Firouzjahi2017Instabilities}. Consequently, the existence of an asymptotic oscillatory regime in the original model should not be interpreted as establishing the existence
of a stable quantum ground state.

 Constructing healthy nonsingular cosmologies within scalar-tensor gravity is itself nontrivial. Nonsingular solutions are strongly constrained within Horndeski theory by no-go results \cite{kobayashi2016nogo}; in an effective-field-theory description, avoiding the associated gradient instability points to operators characteristic of beyond-Horndeski theories \cite{creminelli2016stability}. Complementary constraints arise in single-field matter-bounce scenarios, where suppressing the tensor-to-scalar ratio can conflict with maintaining acceptable primordial non-Gaussianities \cite{quintin2015nogo,li2017extendednogo}. At the same time, generic nondegenerate higher-derivative scalar--tensor theories propagate an additional Ostrogradsky degree of freedom. Degenerate higher-order scalar-tensor (DHOST) theories provide a systematic way to evade this
 pathology through degeneracy conditions that generate additional constraints in phase space \cite{Langlois_2016,langlois2016hamiltonian,ben2016degenerate}. Mimetic theories can themselves be formulated within the DHOST framework and possess an additional local symmetry associated with the non-invertibility of the mimetic transformation \cite{langlois2019mimetic,ben2025dhost}. Nonsingular early-Universe scenarios have also been explored directly within higher-order scalar-tensor and DHOST frameworks, including stable bouncing and emergent/Genesis constructions and studies of their cosmological perturbations \cite{ilyas2020dhost,ilyas2021emergent,zhu2021dhostbounce, volkova2025nonsingular,an2025dhostbounce}. This motivates the central question of the present work: how does the perturbative structure of the bouncing scenario change when its higher-derivative scalar sector is embedded in a quadratic degenerate realization?

 In this work, we construct a quadratic subclass Ia representative whose homogeneous scale-factor evolution can be matched to that of the mimetic bounce of Ref.~\cite{machado2025quantum} after a constant rescaling of the scalar potential. Although the same function $a(t)$ is reproduced, the associated mimetic Lagrange multiplier generally differs. We then perform an ADM and Hamiltonian reduction of the scalar perturbations. The main result is that the reduced canonical scalar mode contains no spatial-gradient contribution, corresponding to a vanishing physical sound speed, $c_s^2=0$,  in agreement with the known vacuum behavior of the \(\alpha_L=0\) mimetic-DHOST sector found in Ref.~\cite{langlois2019mimetic}, which is recovered explicitly here for the subclass Ia bouncing realization. The absence of the gradient term also has an immediate consequence for the initial-state prescription. On the matched bouncing background, the effective frequency squared is independent of the comoving wavenumber and approaches zero from below in the remote contracting phase. Consequently, canonical modes become nonoscillatory, so the asymptotic positive-frequency prescription in Ref.~\cite{machado2025quantum} is not available. Without an additional prescription for the quantum state, the primordial scalar spectrum is not uniquely predicted.

 The paper is organized as follows. In Sec. \ref{sec:iamodel} we introduce the quadratic subclass Ia realization and establish the correspondence between its scale-factor evolution and that of the original mimetic model. In Sec. \ref{sec:scalardynamics} we derive the compact ADM action, analyze the scalar constraints, and obtain the reduced canonical dynamics. In Sec. \ref{sec:vacuumselec} we discuss the immediate consequences of the vanishing spatial-gradient term for the asymptotic initial-state prescription of the matched bounce. We summarize our results in Sec. \ref{sec:concl}.

 \section{Subclass Ia mimetic DHOST realization}
 \label{sec:iamodel}

 \subsection{Quadratic DHOST theories and degeneracy conditions}
 \label{sec:qdhosttheory}

 Horndeski theory constitutes the most general four-dimensional scalar-tensor theory with second-order equations of motion \cite{horndeski1974second}. Generic higher-derivative Lagrangians beyond it propagate an additional degree of freedom: the Hamiltonian is linear in one of the canonical momenta, an instability known as the Ostrogradsky ghost \cite{ostrogradsky1848memoire,woodard2015ostrogradsky}. It was later realized, however, that second-order equations of motion are not a necessary condition for avoiding this extra mode. Beyond-Horndeski theories provided the first examples of scalar-tensor theories with higher-order equations of motion that nevertheless propagate only three degrees of freedom. This observation was generalized in the framework of degenerate higher-order scalar-tensor theories \cite{Langlois_2016,langlois2016hamiltonian}: the kinetic matrix of the quadratic Lagrangian is degenerate, which generates additional constraints in phase space and removes the Ostrogradsky mode. DHOST theories have been systematically classified up to cubic order in second derivatives of the scalar field \cite{Langlois_2016,langlois2016hamiltonian, crisostomi2016horndeski,crisostomi2016extended,
 ben2016degenerate,achour2016degenerate,
 de2016ostrogradsky,langlois2017effective, langlois2019review, lazanu2025dhost}.
 
 For the quadratic theories considered here, the gravitational Lagrangian reads

  \begin{equation}
    \mathcal{L}^{(2)} = f_2(\phi,X)\,R +\sum_{i=1}^{5} a_i(\phi,X)\,L_i^{(2)},
    \label{eq:dhostqaction}
  \end{equation}
 where $f_2(\phi,X)$ and $a_i(\phi,X)$ are free functions. Throughout this work, we adopt the metric signature $(+,-,-,-)$ and define the kinetic invariant

  \begin{equation}
     X \equiv \frac{1}{2}\,g^{\mu\nu}\nabla_\mu\phi\,\nabla_\nu\phi
    \label{eq:kterm},
  \end{equation} 
 together with
 
  \begin{equation}
    \phi_{\mu\nu}\equiv\nabla_\mu\nabla_\nu\phi,
    \qquad \Box\phi\equiv g^{\mu\nu}\nabla_\mu\nabla_\nu\phi.
  \end{equation}
 Natural units $\hbar=c=M_{\rm Pl}=1$ are used. The five quadratic operators are 

  \begin{equation}
   \begin{aligned}
     L_1^{(2)} &= \phi_{\mu\nu}\phi^{\mu\nu},
     &\quad
     L_2^{(2)} &= (\Box\phi)^2, 
     \\[8pt]
      L_3^{(2)} &= (\Box\phi)\,\phi^\mu\phi_{\mu\nu}\phi^\nu,
     &\quad 
     L_4^{(2)} &= \phi^\mu\phi_{\mu\rho}\phi^{\rho\nu}\phi_\nu,
     &\qquad\\[8pt] L_5^{(2)} &= \left(\phi^\mu\phi_{\mu\nu}\phi^\nu\right)^2.
     \end{aligned}
     \label{eq:quadratic_dhost_operators}
  \end{equation}
 
 The degeneracy of the kinetic matrix is encoded in the three conditions
 
  \begin{equation}
    D_0(X)=D_1(X)=D_2(X)=0.
  \end{equation}
 Here $D_0$, $D_1$, and $D_2$ are the coefficient functions entering the polynomial degeneracy condition for the kinetic matrix. Requiring all three to vanish classifies quadratic DHOST theories into three main classes and seven subclasses; their explicit expressions in terms of $f_2$, $f_{2,X}$, and $a_i$ can be found in Refs.~\cite{Langlois_2016,ben2016degenerate}. Ref.~\cite{ben2016degenerate} adopts different conventions for both the kinetic invariant and the metric signature; all degeneracy relations quoted below have been translated to the conventions of Eq.~\eqref{eq:kterm}. In particular, the kinetic invariant used in Ref.~\cite{ben2016degenerate} is related to ours by $X_{\rm Ref.}=2X$. In these conventions, class I is characterized by

  \begin{equation}
    a_1+a_2=0,
    \label{eq:classI}
  \end{equation}
 while subclass Ia additionally satisfies
 
  \begin{equation}
    f_2+2Xa_2 = f_2-2Xa_1 \neq0.
    \label{eq:classIacondition}
  \end{equation}
 For a generic subclass Ia theory, the remaining degeneracy conditions determine $a_4$ and $a_5$ in terms of $f_2$, $a_2$, and $a_3$.

 \subsection{A subclass Ia realization of the mimetic model}
 \label{sec:theory}

 Mimetic gravity can itself be formulated as a degenerate higher-order scalar-tensor theory: the non-invertibility of the mimetic transformation introduces an additional local symmetry and, from the Hamiltonian point of view, an associated first-class constraint \cite{langlois2019mimetic, ben2025dhost}. Our starting point is the higher-derivative mimetic model of Ref.~\cite{machado2025quantum}

  \begin{equation}
    S_{\rm M}=\int d^4x\sqrt{-g}\Big[-\tfrac12 R+\lambda(2X-1)-V_{\rm M}(\phi) +\tfrac{\g}{2}(\Bx\phi)^2\Big],
   \label{eq:SM}
  \end{equation}
 Within subclass Ia, we further restrict attention to the special sector for which the scalar kinetic matrix remains degenerate when the metric is treated as nondynamical. In the notation of Ref.~\cite{ben2016degenerate}, this corresponds to $\mathcal A=0$; in the conventions adopted here, the corresponding relations are
 
  \begin{equation}
    a_3+a_4=0,
    \qquad
    a_5=0,
    \label{eq:specialIa1}
  \end{equation}
 together with
 
  \begin{equation}
    f_{2,X}+a_2+Xa_3=0.
    \label{eq:specialIa2}
  \end{equation}
 Within the ${\cal A}=0$ sector, fixing the Einstein-Hilbert coefficient and retaining the same coefficient $\gamma/2$ of $(\Box\phi)^2$ as in the original model uniquely determines the remaining quadratic coefficients through the subclass Ia degeneracy conditions. To construct the representative considered in this work, we choose
 
  \begin{equation}
    f_2=    - 1/2,
    \qquad
    f_{2,X}=0,
    \qquad
    a_2=\gamma/2,
    \label{eq:f2a2}
  \end{equation}
 where $\gamma$ is a constant. The class-I condition~\eqref{eq:classI} then fixes
 
  \begin{equation}
    a_1=-a_2=-\gamma/2,
    \label{eq:a1model}
  \end{equation}
 and since $f_{2,X}=0$ and $a_2=\gamma/2$, Eqs.~\eqref{eq:specialIa1} and ~\eqref{eq:specialIa2} are provided
 
  \begin{equation}
   a_3=-\frac{\gamma}{2X},
    \qquad
   a_4=\frac{\gamma}{2X}.
   \label{eq:a34}
  \end{equation}
 The particular subclass Ia representative studied here is therefore specified by

  \begin{equation}
   \begin{aligned}
     f_2 &= -\frac{1}{2},
     &\quad a_1 &= -\frac{\gamma}{2},
     &\quad a_2 &= \frac{\gamma}{2},
     \\[3pt] a_3 &= -\frac{\gamma}{2X},
     &\quad a_4 &= \frac{\gamma}{2X},
     &\qquad a_5 &= 0.
   \end{aligned}
   \label{eq:modelcoefficients}
  \end{equation}
 As a consistency check, on the mimetic constraint surface, $X=1/2$, the subclass Ia condition ~\eqref{eq:classIacondition}  becomes 

  \begin{equation}
    f_2+2Xa_2=\frac{\gamma-1}{2}\neq0,
  \end{equation}
 so that this representative requires 
 
  \begin{equation} 
    \gamma\neq1. 
    \label{eq:gammaIa} 
  \end{equation}
 We restrict the discussion to the connected timelike region $X>0$ containing the mimetic constraint surface $X=1/2$ and satisfying $f_2+2Xa_2\neq0$. For the coefficients in Eq.~\eqref{eq:modelcoefficients}, $f_2+2Xa_2=-1/2+\gamma X$. For $\gamma>0$, this quantity vanishes at $X=1/(2\gamma)$, whereas for $\gamma<0$, it has no zero in the timelike region $X>0$. Since $\gamma\neq1$, the zero does not coincide with the mimetic constraint surface. Throughout this work, ``subclass Ia'' therefore refers to the local branch containing $X=1/2$. The restriction $X>0$ is also consistent with the $1/X$ dependence of $a_3$ and $a_4$.

 Substituting these coefficients into Eq.~\eqref{eq:dhostqaction}, the action becomes

  \begin{equation}
   \begin{aligned}
    S_{\rm Ia} = \int d^4x\,\sqrt{-g}\,\Bigg\{& -\frac{1}{2}R +\lambda(2X-1) -V_{\rm Ia}(\phi) 
    \\[3pt] &-\frac{\gamma}{2}\, \phi_{\mu\nu}\phi^{\mu\nu} +\frac{\gamma}{2}\, (\Box\phi)^2
    \\[3pt] &-\frac{\gamma}{2X}\, (\Box\phi)\, \phi^\mu\phi_{\mu\nu}\phi^\nu 
    \\[3pt] &+\frac{\gamma}{2X}\, \phi^\mu\phi_{\mu\rho} \phi^{\rho\nu}\phi_\nu \Bigg\}.
   \end{aligned}
   \label{eq:mgdhostaction}
  \end{equation}

 \subsection{Background matching}
 \label{subsec:backgroundmatching2}

 We compare the homogeneous dynamics of the subclass Ia realization with those of the original mimetic model \eqref{eq:SM}. In unitary gauge, $\phi=t$, one has $X=1/(2N^2)$. The mimetic constraint $2X-1=0$ therefore fixes the positive-lapse branch $N=1$, so that this is not an additional gauge choice. Hence, at the background level,
 
  \begin{equation}
    \phi=t, \quad N=1,
  \end{equation}
 the background equations following from Eq.~\eqref{eq:mgdhostaction} are
 
  \begin{align}
    3H^2 \left( 1 - 3\gamma\right) - 2\lambda_{\rm Ia} -V_{\rm Ia} &=0,
    \label{eq:eofia1}
    \\ (1-\gamma)\left(2\dot H+3H^2\right) &=V_{\rm Ia}.
    \label{eq:eofia2}
  \end{align}
 We denote by $\lambda_{\rm Ia}(t)$ the homogeneous value of the Lagrange multiplier in the subclass Ia realization; the first equation determines $\lambda_{\rm Ia}$ once the background evolution is specified, whereas Eq.~\eqref{eq:eofia2} governs the scale-factor evolution.  

 For comparison, the corresponding equations of the original higher-derivative mimetic model, Eq.~\eqref{eq:SM}, are~\cite{machado2025quantum}

  \begin{align}
    3H^2 -2\lambda_{\rm M} -V_{\rm M} -3\gamma\left(\frac{3}{2}H^2 - \dot H \right) &=0,
   \label{eq:eofm1}
   \\ (2-3\gamma)\left(2\dot H+3H^2\right) &=2V_{\rm M}.
  \label{eq:eofm2}
  \end{align}
 
 Although Eqs.~\eqref{eq:eofia1} and~\eqref{eq:eofm1} determine different Lagrange multipliers, the equations governing the scale-factor evolution can be mapped exactly into one another. Indeed, choosing

  \begin{equation}
    V_{\rm Ia}(\phi) = \frac{2(1-\gamma)}{2-3\gamma}\, V_{\rm M}(\phi),
   \label{eq:rescale}
  \end{equation}
 makes Eqs.~\eqref{eq:eofia2} and~\eqref{eq:eofm2} identical. The mapping in Eq.~\eqref{eq:rescale} is singular at $\gamma=2/3$, while the subclass Ia condition excludes $\gamma=1$. Moreover, in the original mimetic model, the scalar sound speed is $c_{s,\rm M}^2=\gamma/(2-3\gamma)$, which is positive for $0<\gamma<2/3$. For the comparison with Ref.~\cite{machado2025quantum}, we therefore focus on this branch. We emphasize that this restriction is associated with the matching to the original model; the vanishing sound speed derived below for the subclass Ia realization follows instead from its constrained scalar dynamics.

 The difference between the corresponding Lagrange multipliers can be written directly in terms of the common background. Using Eqs.~\eqref{eq:eofia2} and~\eqref{eq:eofm2} to eliminate the potentials from Eqs.~\eqref{eq:eofia1} and~\eqref{eq:eofm1}, respectively, one finds

  \begin{equation}
    \lambda_{\rm Ia}
    =-(1-\gamma)\dot H-3\gamma H^2,
    \qquad
    \lambda_{\rm M}
    =(3\gamma-1)\dot H.
    \label{eq:lambdacomparison}
  \end{equation}
 Within this branch, the two potential families generate a one-to-one correspondence between their scale-factor histories. In particular, any bouncing background obtained in Ref.~\cite{machado2025quantum} can be reproduced in the subclass Ia realization with the same $a(t)$ after the constant rescaling~\eqref{eq:rescale} of the potential.

\begin{figure}[H]
   \centering
   \includegraphics[width=0.4\textwidth]{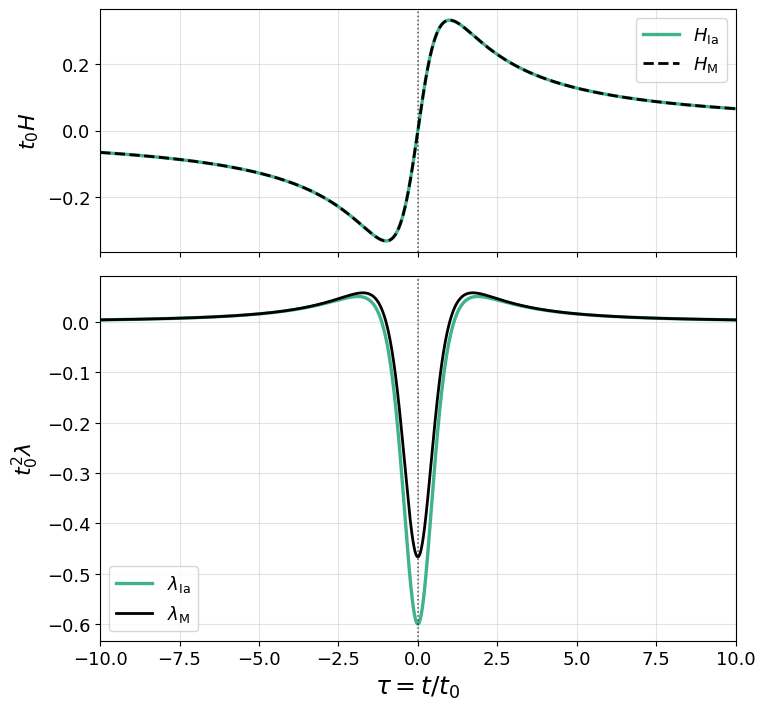}
   \caption{Background matching for the representative matter-like case $\omega=0$ ($r=1/3$), with $\gamma=0.1$. The top panel shows the dimensionless Hubble parameter $t_0H$ for the subclass Ia and original mimetic models; the two curves coincide because the scale-factor histories are matched exactly. The bottom panel shows the corresponding
   dimensionless Lagrange multipliers $t_0^2\lambda_{\rm Ia}$ and $t_0^2\lambda_{\rm M}$, which differ despite the identical homogeneous geometry. The vertical dotted line marks the bounce at $\tau=0$.}
   \label{fig:backgroundmatching2}
 \end{figure}

 To illustrate this correspondence explicitly, we consider the nonsingular background used in Ref.~\cite{machado2025quantum}

  \begin{equation}
    a(\tau)=a_b(1+\tau^2)^r,
    \quad r=\frac{1}{3(1+\omega)},
    \quad \tau=\frac{t}{t_0}.
    \label{eq:bouncer}
  \end{equation}
  Here $\omega$ parametrizes the effective asymptotic equation of state of the matched background, as in Ref.~\cite{machado2025quantum}, rather than representing an independently added perfect-fluid component. The parameter $t_0$ sets the characteristic time scale of the bounce. For the bouncing branch relevant to Ref.~\cite{machado2025quantum}, $\omega>-1/3$, so that $0<r<1/2$. The corresponding Hubble parameter and its time derivative are

  \begin{equation}
    H(\tau) = \frac{2r\tau}{t_0(1+\tau^2)},
    \qquad \dot H(\tau) = \frac{2r(1-\tau^2)}{t_0^2(1+\tau^2)^2}.
    \label{eq:HandHdotexplicit}
  \end{equation}
 In particular, $|H|$ reaches its maximum value $r/t_0$ at $|\tau|=1$, while $\dot H(0)=2r/t_0^2$. Figure~\ref{fig:backgroundmatching2} illustrates the background correspondence for the representative matter-like case $\omega=0$, corresponding to $r=1/3$, with $\gamma=0.1$. The value of $\gamma$ is chosen here only for visual clarity. The Hubble evolution is identical in the two models, whereas the reconstructed Lagrange multipliers differ particularly around the bounce.

 The two models are therefore not identical at the full background level. Although the scale-factor history and, hence, the homogeneous geometry can be matched exactly, the corresponding Lagrange multipliers are generally different. This separation isolates the effect of the subclass Ia structure on the scalar sector without simultaneously changing the bouncing geometry.

\section{Reduced perturbation dynamics}
\label{sec:scalardynamics}

 \subsection{Compact ADM action}
 \label{subsec:compactadm}

 In this section we turn to scalar perturbations around the homogeneous background discussed in Sec. \ref{subsec:backgroundmatching2}. We work in unitary gauge, $\delta\phi=0$, so that the scalar-field fluctuations are encoded in the metric. To determine the scalar constraint structure and identify the reduced dynamical degree of freedom,  we perform a perturbative Arnowitt–Deser–Misner (ADM) decomposition of the metric. The ADM variables are parametrized as 
 
  \begin{equation}
    N=1+\alpha, \qquad
    N_i=\partial_i\beta, \qquad
    h_{ij}=a^2(t)e^{2\zeta}\delta_{ij},
    \label{eq:scalar_perturbations}
  \end{equation}
 where $\alpha$ and $\beta$ denote the lapse and shift perturbations, respectively, while $\zeta$ is the curvature perturbation. The second derivatives of $\phi$ appearing in Eq.~\eqref{eq:mgdhostaction} can be expressed directly in terms of ADM quantities. We define

  \begin{equation}
    {\cal B}\equiv \dot N-N^iD_iN,
    \qquad
    s_i\equiv D_i\ln N,
   \label{eq:B_ai_def}
  \end{equation}
 where $s_i$ is the spatial lapse-gradient vector. In unitary gauge one obtains
  
  \begin{align}
    &\Box\phi = KN^{-1} - {\cal B}N^{-3},
    \label{eq:boxphi_adm_short}
    \\ &\phi_{\mu\nu}\phi^{\mu\nu} =N^{-2}\left(K_{ij}K^{ij}-2s_is^i\right) +{\cal B}^2N^{-6},
    \label{eq:phimunu_adm_short}
  \end{align}
 The remaining contractions entering the DHOST sector give

  \begin{equation}
    \phi^\mu\phi_{\mu\nu}\phi^\nu =-{\cal B} N^{-5}, \quad
    \phi^\mu\phi_{\mu\rho}\phi^{\rho\nu}\phi_\nu ={\cal B}^2N^{-8}-s_is^iN^{-4}.
    \label{eq:JK_adm_short}
  \end{equation}
 In unitary gauge, $X=1/(2N^2)$, so that the coefficients in Eq.~\eqref{eq:modelcoefficients} become $a_3=-\gamma N^2$ and $a_4=\gamma N^2$. Substituting the contractions above into the DHOST sector, all terms involving ${\cal B}$ and $s_i$ cancel, leaving
 
  \begin{equation}
   \begin{aligned} 
    {\cal L}_{\rm D} ={}& [\gamma /(2N^2)] \left( K^2-K_{ij}K^{ij}
    \right).
   \end{aligned}
  \label{eq:dhostadmcompact}
  \end{equation}
 The complete action can therefore be written as

  \begin{equation}
   \begin{aligned}
    S=\int dt\,d^3x\,N\sqrt h\, \Bigg[ &-\frac12\,{}^{(3)}R +\lambda\left(\frac{1}{N^2}-1\right) -V(\phi)
    \\ &+\frac12  \left(1 - \frac{\gamma}{N^2} \right) \left(K_{ij}K^{ij} -K^2 \right) \Bigg].
   \end{aligned}
   \label{eq:fulladmcompact}
  \end{equation}
 At $N=1$ the extrinsic-curvature sector reduces to $(1-\gamma)(K_{ij}K^{ij}-K^2)/2$. The higher-derivative sector therefore changes only the coefficient of the standard extrinsic-curvature combination, while all terms involving the spatial lapse gradient and ${\cal B}$ cancel identically.

 \subsection{Reduced action and constraints}
 \label{subsec:reducedaction}

 Imposing the mimetic constraint and the background equations reduces the scalar quadratic action as follows. Writing the Lagrange multiplier as

  \begin{equation}
    \lambda(t,\mathbf{x}) =\lambda_{\rm Ia}(t)+\delta\lambda(t,\mathbf{x}),
  \end{equation}
 the variation with respect to $\delta\lambda$ gives
 
  \begin{equation}
    \frac{\delta S^{(2)}}{\delta(\delta\lambda)} = -2a^3\alpha = 0.
  \end{equation}
 Hence,
 
  \begin{equation}
    \alpha=0.
    \label{eq:alphazero}
  \end{equation}
 This constraint holds mode by mode, including the homogeneous sector, so that no independent lapse perturbation survives. After imposing Eq.~\eqref{eq:alphazero}, the scalar quadratic action takes the form

  \begin{equation}
   \begin{aligned}
     S^{(2)}\big|_{\alpha=0} = \int dt \,d^3x\,a^3 \Bigg[ & -3(1-\gamma)\dot\zeta^2 -18(1-\gamma)H\zeta\dot\zeta
     \\ & -\left( \frac{27}{2}(1-\gamma)H^2 +\frac92 V_{\rm Ia} \right)\zeta^2
     \\ & -\frac{(\partial\zeta)^2}{a^2} +2(1-\gamma) \frac{\dot\zeta}{a^2}\partial^2\beta \Bigg].
    \end{aligned}
   \label{eq:quadraticalphazero}
  \end{equation}
 The term proportional to $H\zeta\dot\zeta$ can be integrated by parts in time. Up to a boundary term,
 
  \begin{equation}
   \begin{aligned}
     -18(1-\gamma)\int dta^3H\zeta\dot\zeta &= 9(1-\gamma)\int dta^3 \left(\dot H+3H^2\right)\zeta^2 .
   \end{aligned}
\end{equation}
 The resulting coefficient of $\zeta^2$ is
 
  \begin{equation}
   (9/2) \left[ (1-\gamma) \left( 2\dot H+3H^2 \right) - V_{\rm Ia} \right],
  \end{equation}
 which vanishes by the background equation

  \begin{equation}
    (1-\gamma) \left( 2\dot H+3H^2\right) = V_{\rm Ia}.
    \label{eq:backgroundscalarreduction}
  \end{equation}
 The reduced scalar quadratic action is therefore

  \begin{equation}
   \begin{aligned}
    S_{\rm red}^{(2)} =\int dt\,d^3x\,a^3 \Bigg[ &-3(1-\gamma)\dot\zeta^2-\frac{(\partial\zeta)^2}{a^2}
    \\ & +2(1-\gamma) \frac{\dot\zeta}{a^2}\partial^2\beta \Bigg].
   \end{aligned}
   \label{eq:scalarreducedaction}
  \end{equation}
 The shift perturbation $\beta$ enters Eq.~\eqref{eq:scalarreducedaction} without time derivatives and acts as a Lagrange multiplier at this stage. Varying with respect to $\beta$ gives
 
  \begin{equation}
     \delta_\beta S_{\rm red}^{(2)} = 2(1-\gamma) \int dt\,d^3x\, a\,\partial^2\dot\zeta\,\delta\beta,
  \end{equation}
 where spatial boundary terms have been discarded. The momentum constraint is thus

  \begin{equation}
    (1-\gamma)\partial^2\dot\zeta=0.
    \label{eq:zetaconstraintposition}
  \end{equation}
  Introducing the Fourier decomposition

  \begin{equation}
    \zeta(t,\mathbf{x}) = \int\frac{d^3k}{(2\pi)^3} \zeta_{\mathbf{k}}(t)
    e^{i\mathbf{k}\cdot\mathbf{x}},
  \end{equation}
  
  \begin{equation}
    \beta(t,\mathbf{x}) = \int\frac{d^3k}{(2\pi)^3} \beta_{\mathbf{k}}(t) e^{i\mathbf{k}\cdot\mathbf{x}}.
  \end{equation}
 Since $\zeta$ and $\beta$ are real fields, their Fourier coefficients satisfy the reality conditions
 $ \zeta_{-\mathbf{k}} = \zeta_{\mathbf{k}}^{*}, \beta_{-\mathbf{k}} = \beta_{\mathbf{k}}^{*}$. Equation~\eqref{eq:scalarreducedaction} then becomes 

  \begin{equation}
   \begin{aligned}
    S_{\rm red}^{(2)} = \int dt \int\frac{d^3k}{(2\pi)^3} \Bigg[ & -3(1-\gamma)a^3 \dot\zeta_{\mathbf{k}} \dot\zeta_{-\mathbf{k}} -ak^2 \zeta_{\mathbf{k}} \zeta_{-\mathbf{k}}
    \\ & -(1-\gamma)ak^2 \left(\dot\zeta_{\mathbf{k}}\beta_{-\mathbf{k}} + \dot\zeta_{-\mathbf{k}}\beta_{\mathbf{k}} \right)\Bigg],
   \end{aligned}
   \label{eq:scalarreducedfourier}
  \end{equation}
 where $k\equiv|\mathbf{k}|$. For nonzero Fourier modes and $\gamma\neq1$, this implies
 
  \begin{equation}
    \dot\zeta_{\mathbf{k}}=0,
    \qquad
    k\neq0.
    \label{eq:zetadotzero}
  \end{equation}
 The homogeneous mode $k=0$ is not constrained by the momentum constraint. It corresponds to a homogeneous deformation of the background scale factor and is not part of the inhomogeneous perturbative sector considered here. In what follows, the scalar reduction and canonical normalization therefore refer to $k\neq0$. The constraint~\eqref{eq:zetadotzero} must be imposed before interpreting the kinetic and gradient terms in Eq.~\eqref{eq:scalarreducedaction}. In particular, their coefficient ratio does not define the physical scalar sound speed because $\zeta$ is not an unconstrained propagating variable. Combining Eq.~\eqref{eq:zetadotzero} with the equation obtained by varying the reduced action with respect to $\zeta$,
 
  \begin{equation}
   \begin{aligned}
     0={}& 3(1-\gamma)\ddot\zeta+9(1-\gamma)H\dot\zeta+\frac{\partial^2\zeta}{a^2}
     \\&-(1-\gamma) \frac{\partial^2\dot\beta}{a^2} - (1-\gamma)H \frac{\partial^2\beta}{a^2},
   \end{aligned}
   \label{eq:zetaequationreduced}
  \end{equation}
 one obtains, for $k\neq0$,

  \begin{equation}
    \zeta_{\mathbf{k}} = (1-\gamma)\left( \dot\beta_{\mathbf{k}}   +H\beta_{\mathbf{k}} \right).
    \label{eq:zetabetarelation}
 \end{equation}
 Thus, although the curvature perturbation is conserved, the scalar sector still contains nontrivial dynamics encoded in the shift, or equivalently, in the momentum conjugate to $\zeta$.

 \subsection{Hamiltonian reduction and canonical scalar dynamics}
 \label{subsec:hamiltonianreduction}

 The constrained structure of the scalar sector becomes more transparent in the Hamiltonian formulation. To avoid double counting the Fourier degrees of freedom, we first consider one independent real mode and suppress the vector notation on its wavenumber. After reducing mode by mode, we restore the full Fourier representation using the reality conditions. The quadratic Lagrangian for this real mode is 
 
  \begin{equation} 
    L_k^{(2)} = -3(1-\gamma) a^3\dot\zeta_k^2 -ak^2\zeta_k^2 -2(1-\gamma)ak^2\dot\zeta_k\beta_k. 
   \label{eq:realmodelagrangian} 
  \end{equation}
 The momentum conjugate to $\zeta_k$ is 
 
  \begin{equation} 
   \Pi_{\zeta,k} = -6(1-\gamma)a^3\dot\zeta_k -2(1-\gamma)ak^2\beta_k, \label{eq:Pizeta} 
  \end{equation} 
 and the corresponding Hamiltonian is

  \begin{equation} 
    {\cal H}_k = -\frac{ \left[ \Pi_{\zeta,k} +2(1-\gamma)ak^2\beta_k \right]^2 } {12(1-\gamma)a^3} + ak^2\zeta_k^2. 
    \label{eq:Hamiltonianzeta} 
  \end{equation}

 The reduced Lagrangian contains no time derivative of $\beta_k$, so its conjugate momentum vanishes identically,
 
  \begin{equation} 
    \Pi_{\beta,k}=0, 
  \end{equation} 
 defining a primary constraint. Requiring its preservation under Hamiltonian evolution, or equivalently varying Eq.~\eqref{eq:Hamiltonianzeta} with respect to $\beta_k$, gives the secondary constraint 
 
  \begin{equation} 
    \Pi_{\zeta,k} = -2(1-\gamma)ak^2\beta_k. 
    \label{eq:Pibetarelation} 
  \end{equation} 
 Defining the two constraints as
 
  \begin{equation}
    \Phi_1\equiv\Pi_{\beta,k}\approx0,
    \qquad
    \Phi_2\equiv \Pi_{\zeta,k} +2(1-\gamma)ak^2\beta_k \approx0,
  \end{equation}
  their Poisson bracket is
 
  \begin{equation}
    \{\Phi_1,\Phi_2\} = -2(1-\gamma)ak^2.
  \end{equation}

 For $k\neq0$ and $\gamma\neq1$, the constraint matrix is therefore nondegenerate, so that $\Phi_1$ and $\Phi_2$ form a second-class pair. Preservation of the secondary constraint fixes the Lagrange multiplier associated with $\Pi_{\beta,k}$ and generates no further constraint. Starting from the two canonical pairs $(\zeta_k,\Pi_{\zeta,k})$ and $(\beta_k,\Pi_{\beta,k})$, the two second-class constraints leave one physical scalar degree of freedom per nonzero Fourier mode,
 
  \begin{equation}
    N_{\rm dof} = \frac{4-2}{2} = 1.
  \end{equation} 
 Notice also that the negative quadratic contribution in Eq.~\eqref{eq:Hamiltonianzeta} is proportional to $\Phi_2^2$ and therefore vanishes on the constraint surface. It does not correspond to an independent negative-energy degree of freedom.

 Together with Eq.~\eqref{eq:Pizeta}, this relation immediately implies 
 
  \begin{equation} 
    \dot\zeta_k=0, 
  \end{equation} 
 in agreement with the momentum constraint derived directly from the Lagrangian formulation.  After imposing Eq.~\eqref{eq:Pibetarelation}, the first-order action reduces to
 
  \begin{equation}
    S_k^{(2)} = \int dt \left( \Pi_{\zeta,k}\dot\zeta_k - ak^2\zeta_k^2 \right).
    \label{eq:firstorderreduced}
  \end{equation}
 Integrating the first term by parts and discarding the boundary term, one finds 
 
  \begin{equation} 
    S_k^{(2)} = \int dt \left( -\dot\Pi_{\zeta,k}\zeta_k - ak^2\zeta_k^2 \right). 
  \end{equation} 
 Variation with respect to $\zeta_k$ is given  
 
  \begin{equation} 
    \zeta_k = -\frac{\dot\Pi_{\zeta,k}}{2ak^2}. 
    \label{eq:zetaintermsPi} 
  \end{equation} 
 Restoring the full Fourier representation with $\Pi_{\zeta,-\mathbf{k}} = \Pi_{\zeta,\mathbf{k}}^*$, the reduced action becomes:
  
  \begin{equation}
    S_{\Pi}^{(2)} = \frac14 \int dt \int\frac{d^3k}{(2\pi)^3}
    \frac{ \dot\Pi_{\zeta,\mathbf k} \dot\Pi_{\zeta,-\mathbf k}}
    {ak^2}.
   \label{eq:Pireducedaction}
  \end{equation}
 Consistent with the degree-of-freedom count above, Eq.~\eqref{eq:Pireducedaction} shows explicitly that the scalar mode has not disappeared when $\dot\zeta_{\mathbf{k}}=0$. Rather, after imposing the second-class constraints, its nontrivial dynamics are encoded in the momentum conjugate to $\zeta$, or equivalently, in the shift perturbation. Using Eq.~\eqref{eq:Pibetarelation}, 
 
  \begin{equation} 
    \Pi_{\zeta,\mathbf{k}} = -2(1-\gamma)ak^2\beta_{\mathbf{k}},
  \end{equation} 
 Eq.~\eqref{eq:Pireducedaction} can equivalently be written as 
 
  \begin{equation} 
    \begin{aligned} 
     S_{\beta}^{(2)} = (1-\gamma)^2 \int dt \int\frac{d^3k}{(2\pi)^3}\, ak^2 & \left( \dot\beta_{\mathbf{k}} +H\beta_{\mathbf{k}} \right) \\ & \left( \dot\beta_{-\mathbf{k}} +H\beta_{-\mathbf{k}} \right). \end{aligned} 
     \label{eq:betareducedaction} 
  \end{equation} 
 It is convenient to introduce
 
  \begin{equation}
    u_{\mathbf k}\equiv a\beta_{\mathbf k},
    \label{eq:udef}
  \end{equation}
 so that
 
  \begin{equation}
    \dot\beta_{\mathbf k} +H\beta_{\mathbf k} = \frac{\dot u_{\mathbf k}}{a}.
  \end{equation}
 The action becomes 
 
  \begin{equation} 
    S_{\beta}^{(2)} = (1-\gamma)^2 \int dt \int\frac{d^3k}{(2\pi)^3} \frac{k^2}{a} \dot u_{\mathbf{k}} \dot u_{-\mathbf{k}}. 
  \end{equation}
 For the inhomogeneous sector $k\neq0$ considered here, a canonically normalized variable is therefore

 \begin{equation} 
   v_{\mathbf{k}} \equiv \frac{ \sqrt{2}\,|1-\gamma|\,k } {\sqrt a} u_{\mathbf{k}} = \sqrt{2}\,|1-\gamma|\,k\sqrt a\, \beta_{\mathbf{k}}. \label{eq:vdef} 
  \end{equation}
 The homogeneous mode $k=0$ is excluded from this canonical redefinition, consistent with the discussion following Eq.~\eqref{eq:zetadotzero}. In terms of $v_{\mathbf{k}}$, and after discarding a total time derivative, the scalar action takes the canonical form

  \begin{equation} 
   \begin{aligned} 
    S_v^{(2)} = \frac12 \int dt \int\frac{d^3k}{(2\pi)^3} \Bigg[ & \dot v_{\mathbf{k}} \dot v_{-\mathbf{k}} + \left( \frac{H^2}{4} -\frac{\dot H}{2} \right) v_{\mathbf{k}} v_{-\mathbf{k}} \Bigg]. \end{aligned} 
    \label{eq:canonicalscalaraction} 
  \end{equation}
 Importantly, the reduced canonical scalar mode has a positive kinetic term. Thus, the reduced quadratic scalar action does not exhibit the ordinary kinetic ghost identified in the minimal higher-derivative model of Ref.~\cite{machado2025quantum}. This conclusion applies to the quadratic inhomogeneous scalar sector with $k\neq 0$. It does not establish nonlinear stability or determine the strong-coupling scale, which requires analysis of higher-order interactions. The homogeneous $k=0$ mode, excluded from the canonical transformation ~\eqref{eq:vdef}, must also be treated separately.
 
 The corresponding mode equation is
 
  \begin{equation}
    \ddot v_{\mathbf k} + \left( \frac{\dot H}{2} -\frac{H^2}{4}
    \right) v_{\mathbf k} =0.
    \label{eq:canonicalmodeequation}
  \end{equation}
 We define the effective frequency as
 
  \begin{equation} 
   \Omega^2(t) \equiv \frac{\dot H}{2} -\frac{H^2}{4}. \label{eq:effectivefrequency} 
  \end{equation} 
 Unlike the usual canonical scalar mode equation, $\Omega^2$ contains no $\propto k^2/a^2$ contribution. Thus, increasing the wavenumber does not generate a short-wavelength propagation regime, and the physical scalar sound speed is 
 
  \begin{equation} 
   c_s^2=0. 
   \label{eq:cszero} 
  \end{equation}
 This result is independent of the explicit form of the potential: $V(\phi)$ affects the scalar dynamics through the background functions $H$ and $\dot H$, but does not generate a spatial propagation term.  The absence of the spatial-gradient term is the central result of the scalar reduction. Our explicit ADM and Hamiltonian analysis therefore reproduces, within the subclass Ia realization considered here, the vanishing scalar sound speed found for mimetic DHOST theories in the absence of matter \cite{arroja2016cosmological,langlois2019mimetic}.

 The analysis above refers to the pure mimetic-DHOST scalar sector without an additional matter component. This restriction is important because coupling external matter can modify the scalar constraint structure.  In particular, Ref.~\cite{langlois2019mimetic} showed that including a $k$-essence matter field introduces additional mixing terms into the quadratic scalar action, so that the reduced dynamics differ from the vacuum case. Consequently, the result $c_s^2=0$ obtained here, as well as the initial-state discussion below, should be understood as properties of the pure mimetic sector considered in this work. Whether the same asymptotic behavior persists in a matter-coupled extension requires a separate analysis.

\subsection{Tensor sector}
\label{subsec:tensorsector}

 For completeness, the tensor sector follows directly from the ADM action~\eqref{eq:fulladmcompact}. Introducing a transverse and traceless perturbation $h_{ij}^{T}$,
 
  \begin{equation}
    \partial_i h_{ij}^{T}=0,
    \quad
    h^{T\,i}_{\ \ i}=0,
  \end{equation}
 the quadratic tensor action is
 
  \begin{equation}
    S_T^{(2)} = \frac{1}{8} \int dt\,d^3x\,a^3 \left[ (1-\gamma)\dot h_{ij}^{T}\dot h_{ij}^{T} - \frac{(\partial_k h_{ij}^{T})^2}{a^2} \right].
    \label{eq:tensoraction}
  \end{equation}
 It can be written in standard form
  
  \begin{equation}
    S_T^{(2)} = \frac{1}{8} \int dt\,d^3x\,a^3 \left[ {\cal G}_T\,\dot h_{ij}^{T}\dot h_{ij}^{T} - {\cal F}_T\,
        \frac{(\partial_k h_{ij}^{T})^2}{a^2} \right],
  \end{equation}
 with
 
  \begin{equation}
    {\cal G}_T=1-\gamma,
    \qquad
    {\cal F}_T=1,
    \qquad
    c_T^2 = \frac{{\cal F}_T}{{\cal G}_T} = \frac{1}{1-\gamma}.
    \label{eq:tensorspeed}
  \end{equation} 
 The factor ${\cal G}_T=1-\gamma$ therefore represents the rescaled tensor kinetic normalization or, equivalently, the effective tensor Planck-mass squared in the units adopted here. Since ${\cal G}_T$ is constant, each tensor polarization obeys

  \begin{equation}
    \ddot h_{\lambda,k} + 3H\dot h_{\lambda,k} +c_T^2\frac{k^2}{a^2}h_{\lambda,k}=0.
  \end{equation}
 By contrast, in the original higher-derivative model \eqref{eq:SM}, the additional operator reduces in unitary gauge to $(\Box\phi)^2=K^2$. For a transverse-traceless tensor perturbation in the exponential parametrization, $K=3H$; this operator does not modify the quadratic tensor propagation. The original model therefore has $c_{T,\rm M}^2=1$, whereas the subclass Ia realization has $c_T^2=1/(1-\gamma)$. Thus, the two constructions also differ in their tensor propagation despite sharing the same scale-factor history.
 
 On the branch $0<\gamma<2/3$ considered here, ${\cal G}_T>0$ and ${\cal F}_T>0$; the tensor sector is free of ghost and gradient instabilities at quadratic order. However, the same interval implies 
 
  \begin{equation}
    1<c_T^2<3,
  \end{equation}
 so tensor modes propagate superluminally with respect to the metric light cone in this representative. This property is distinct from the ghost and gradient stability conditions above. Since the present construction is intended as an effective description of the early Universe, we do not assume that the constant-$\gamma$ tensor propagation speed persists to late cosmological times; establishing a viable late-time extension lies beyond the scope of the present work \footnote{If the constant-$\gamma$ action were extrapolated unchanged to late cosmological times, the resulting $c_T\neq1$ would be incompatible with the multimessenger bound from GW170817/GRB170817A~\cite{abbott2017gwgrb}.}. Unlike the scalar sector, tensor modes, therefore, retain a spatial propagation term.

 \section{Implications for the asymptotic initial state}
 \label{sec:vacuumselec}
 
 The specification of primordial fluctuations in bouncing cosmologies is generally tied to the asymptotic regime and the matter content of the model, and the subsequent transfer of perturbations through the bounce has been studied in a variety of nonsingular scenarios \cite{cai2008quintomperturbations, cai2009metricperturbations,cai2015lcdmbounce}. Alternative prescriptions in which primordial fluctuations have a non-vacuum origin, such as thermal fluctuations generated during contraction, have also been considered \cite{cai2009thermal}. In the present model, however, the canonical action derived in the previous section already contains the essential result for the scalar perturbations: the mode equation has no spatial-gradient contribution and hence no frequency scale that grows with the comoving wavenumber. The large-$k$ limit therefore does not generate the short-wavelength adiabatic hierarchy that is normally used to select a preferred positive-frequency state \cite{bianchi2026adiabatic}. 
 
\begin{figure}[H]
   \centering
   \includegraphics[width=0.4\textwidth]{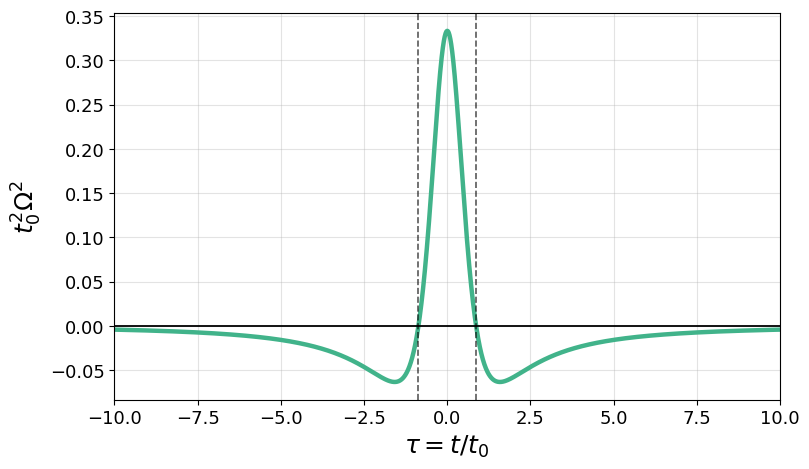}
   \caption{Dimensionless effective frequency $t_0^2\Omega^2$ for the representative matter-like bounce, $\omega=0$ ($r=1/3$). The effective frequency squared is positive only in a finite interval around the bounce, vanishes at $|\tau|=\sqrt{3}/2$, and approaches zero from below in the asymptotic contracting and expanding phases.}
   \label{fig:effectivefrequency}
 \end{figure}

 We now examine this consequence within the pure mimetic sector for the bouncing background matched in Sec. \ref{subsec:backgroundmatching2}. For the matched background~\eqref{eq:bouncer}, substitution of Eq.~\eqref{eq:HandHdotexplicit} into Eq.~\eqref{eq:effectivefrequency} gives

  \begin{equation}
    \Omega^2(\tau) = \frac{r\left[1-(1+r)\tau^2\right]}{t_0^2(1+\tau^2)^2}.
    \label{eq:omegaexplicit}
  \end{equation}
 In the representative matter-like background used in Fig.~\ref{fig:backgroundmatching2}, $r=1/3$, the effective frequency squared is positive only in a finite interval around the bounce. It vanishes at $|\tau|=1/\sqrt{1+r}=\sqrt{3}/2$ and becomes negative outside this interval, as shown in Fig.~\ref{fig:effectivefrequency}.

 In the remote contracting phase,
 
  \begin{equation}
    \Omega^2(t)
    \simeq
    -\frac{r(1+r)}{t^2}
    \longrightarrow0^-,
    \qquad
    t\rightarrow-\infty.
    \label{eq:omegaasymptotic}
  \end{equation} 
 The negative sign of $\Omega^2$ should not be confused with a gradient instability, since $\Omega^2$ is $k$-independent and the asymptotic solutions are power laws rather than $k$-enhanced exponentials. The asymptotic solutions are therefore nonoscillatory. In particular, the matched subclass Ia model has no asymptotic positive-frequency regime from which the initial quantum state could be selected.

 In Fourier space, the canonical scalar mode of Ref.~\cite{machado2025quantum} can be written, up to an irrelevant phase associated with the spatial derivative, as

  \begin{equation}
    v_k^{\rm M} = \frac{\sqrt{\gamma}}{c_s}\,
    k\,\delta\phi_k.
    \label{eq:canonicalprevious}
  \end{equation}
 Its mode equation is

  \begin{equation}
    {v_k^{\rm M}}'' + \left[ c_s^2k^2 +\frac{a''}{a}-2\left(\frac{a'}{a} \right)^2 \right]v_k^{\rm M}=0,
    \quad c_s^2=\frac{\gamma}{2-3\gamma},
    \label{eq:previousquantode}
  \end{equation}
 where primes denote derivatives with respect to conformal time, $d\eta=dt/a$. Equivalently, defining the background pump field $z_{\rm M}\equiv a^{-1}$, Eq.~\eqref{eq:previousquantode} takes the Mukhanov-type form

  \begin{equation}
    {v_k^{\rm M}}'' + \left( c_s^2k^2 - \frac{z_{\rm M}''}{z_{\rm M}} \right)v_k^{\rm M} =0.
  \end{equation}
 On the branch $0<\gamma<2/3$ specified in Sec.~\ref{subsec:backgroundmatching2}, one has $c_s^2>0$. Moreover, since $r<1/2$, the conformal time satisfies $\eta\rightarrow-\infty$ in the remote contracting phase. The condition $|c_sk\eta|\gg1$ then reduces Eq.~\eqref{eq:previousquantode} to ${v_k^{\rm M}}''+c_s^2k^2v_k^{\rm M}\simeq0$, providing the formal positive-frequency prescription used in Ref.~\cite{machado2025quantum} to specify the initial mode functions. No analogous asymptotic prescription exists for Eq.~\eqref{eq:canonicalmodeequation}. The scale-invariant result obtained in Ref.~\cite{machado2025quantum} relied, in particular, on the $k$-dependence fixed by this asymptotic vacuum prescription, together with the subsequent mode evolution. Once this prescription is absent, no analogous spectral $k$-dependence is dynamically selected in the subclass Ia model.

 The absence of such a prescription should not be confused with an obstruction to writing canonically normalized mode functions. The general solution of Eq.~\eqref{eq:canonicalmodeequation} is
 
  \begin{equation}
    v_{\mathbf k}(t) = \frac{1}{\sqrt{a(t)}} \left[C_1(\mathbf k) + C_2(\mathbf k)\int_{t_*}^{t} a(t')\,dt' \right],
    \label{eq:vexactsolution}
  \end{equation}
 and the canonical Wronskian normalization imposes
 
  \begin{equation}
    C_1(k)C_2^*(k)-C_1^*(k)C_2(k)=i.
    \label{eq:WronskianC1C2}
  \end{equation}
 The reference time $t_*$ is arbitrary, since a change in its value can be absorbed into a redefinition of $C_1(\mathbf k)$. For the background~\eqref{eq:bouncer}, $a(t)\propto |t|^{2r}$ as $t\rightarrow-\infty$. Hence, the two independent contributions to Eq.~\eqref{eq:vexactsolution} behave as $|t|^{-r}$ and $|t|^{r+1}$, respectively. If boundedness of the canonical variable in the remote past is imposed as an additional condition, it requires $C_2(k)=0$, which is incompatible with the Wronskian normalization \eqref{eq:WronskianC1C2}. Thus, canonical normalization and asymptotic boundedness of $v_k$ cannot be imposed simultaneously. 
 This tension does not, by itself, establish a strong-coupling regime, since determining the nonlinear strong-coupling scale requires an analysis of higher-order interactions beyond the quadratic action.

 Using Eqs.~\eqref{eq:zetabetarelation}, \eqref{eq:udef}, and~\eqref{eq:vdef}, the curvature perturbation associated with the general solution~\eqref{eq:vexactsolution} is

  \begin{equation}
    \zeta_{\mathbf k} = {\rm sgn}(1-\gamma)\, \frac{C_2(\mathbf k)}{\sqrt{2}\,k},
    \label{eq:zetaintermsC2}
  \end{equation}

 for $k\neq0$. The corresponding dimensionless power spectrum is

  \begin{equation}
    {\cal P}_{\zeta}(k)  = \frac{k^3}{2\pi^2}|\zeta_{\mathbf k}|^2 = \frac{k}{4\pi^2}|C_2(k)|^2.
    \label{eq:PzetaC2}
  \end{equation}

 Thus the spectral shape is controlled directly by the $k$-dependence of $C_2(k)$. For example, a $k$-independent choice of $C_2$ would give a blue spectrum, ${\cal P}_{\zeta}\propto k$, whereas scale invariance would require $|C_2(k)|^2\propto k^{-1}$. Since the asymptotic dynamics does not select $C_2(k)$, neither behavior is preferred without an additional initial-state prescription. The absence of an asymptotic state-selection criterion therefore does not single out a particular spectral shape; rather, the $k$-dependence of $C_2(k)$ remains undetermined within the pure mimetic sector. Unlike the spectrum of the Newtonian potential $\psi=\dot{\delta\phi}$ studied in Ref.~\cite{machado2025quantum}, Eq.~\eqref{eq:PzetaC2} describes the curvature perturbation $\zeta$. Our comparison therefore concerns the asymptotic initial-state prescriptions rather than a direct identification of the two power spectra.

\section{Conclusions}
\label{sec:concl}

 In this paper, we investigated the scalar sector of a quadratic subclass Ia mimetic DHOST realization of the nonsingular bouncing cosmology studied in Ref.~\cite{machado2025quantum}. We showed that a constant rescaling of the scalar potential allows the two constructions to reproduce the same scale-factor evolution, although their mimetic Lagrange multipliers generally differ.

 The main result follows from the ADM and Hamiltonian reduction of the scalar perturbations. The Hamiltonian analysis shows that the two second-class constraints leave one physical scalar degree of freedom for $k\neq0$. After reduction, its canonical action has a positive kinetic term but contains no spatial-gradient contribution, leading to a vanishing physical sound speed, $c_s^2=0$. The curvature perturbation is conserved for nonzero Fourier modes, while the remaining scalar dynamics are carried by the conjugate momentum, or equivalently by the shift sector. This explicit reduction reproduces the known mimetic-DHOST behavior within the subclass Ia bouncing realization considered here. 

 An immediate consequence is that the effective frequency squared does not acquire a wavenumber-dependent short-wavelength contribution. For the matched bounce, it approaches zero from below in the remote contracting phase, so the asymptotic oscillatory regime used in Ref.~\cite{machado2025quantum} to specify the initial quantum state is absent. Canonical normalization can still be imposed through the Wronskian condition, but it does not select a unique state. Without an additional initial-state prescription, the primordial scalar spectrum is therefore not uniquely predicted. The comparison shows that preserving the same nonsingular scale-factor evolution does not imply preserving the perturbative structure or the initial-state prescription of the original model.

 For completeness, the tensor sector has ${\cal G}_T=1-\gamma$ and $c_T^2=1/(1-\gamma)$. On the branch $0<\gamma<2/3$, it is free of ghost and gradient instabilities at quadratic order, although tensor modes propagate superluminally with respect to the metric light cone. Unlike the scalar sector, tensor modes retain a nonvanishing spatial propagation term.

 A complete phenomenological assessment would require an additional prescription fixing the scalar initial state, together with a calculation of the primordial tensor spectrum and a matter-coupled extension. This is particularly relevant because bouncing cosmologies can leave observable signatures in primordial or relic gravitational waves and, through the small-scale scalar spectrum, in primordial black holes and scalar-induced gravitational waves \cite{Chen2017pbh,papanikolaou2024pbh,zhu2026relicgw}. We leave such an observational analysis for future work.

\begin{acknowledgments}
 ILM expresses her gratitude to CAPES for supporting her PhD project and thanks the International Initiative of Particle Cosmology at the University of Science and Technology of China (COSPAI@USTC) for their kind hospitality during the development of this work. ILM is also grateful to Amara Ilyas, Nelson Pinto-Neto, Sanket Dave, and Zhu Mian for their valuable comments. This work is supported in part by the National Key R\&D Program of China (2021YFC2203100), by the NSFC (12433002, 12261131497, W2611007), by CAS young interdisciplinary innovation team (JCTD-2022-20), by 111 Project (B23042), by CSC Innovation Talent Funds, by USTC Fellowship for International Cooperation, and by USTC Research Funds of the Double First-Class Initiative. 
\end{acknowledgments}

\bibliography{ref.bib}

\end{document}